\documentclass[prd, twocolumn, nofootinbib, tightenlines,preprintnumbers,notitlepage,longbibliography,superscriptaddress]{revtex4-1}

\usepackage{graphicx}
\usepackage{hhline}
\usepackage{amsmath}
\usepackage{amsfonts}
\usepackage[colorlinks=true,citecolor=blue,urlcolor=cyan,linkcolor=black]{hyperref}
\usepackage{tablefootnote}
\usepackage{multirow}
\usepackage{relsize}

\usepackage{aas_macros}

\usepackage{setspace}
\usepackage[hang]{footmisc} 
\usepackage{scalerel,stackengine}
\stackMath
\newcommand\what[1]{%
\savestack{\tmpbox}{\stretchto{%
  \scaleto{%
    \scalerel*[\widthof{\ensuremath{#1}}]{\kern-.6pt\bigwedge\kern-.6pt}%
    {\rule[-\textheight/2]{1ex}{\textheight}}
  }{\textheight}%
}{0.5ex}}%
\stackon[1pt]{#1}{\tmpbox}%
}

\usepackage{dcolumn}
\usepackage{bm}
\usepackage[normalem]{ulem}

\usepackage{xcolor}

\usepackage[export]{adjustbox}

\usepackage{suffix}
\usepackage{mathtools}

\newcommand{\fnlloc}{f_{NL}^{\rm{loc}}}

\begin{document}

\title{The kSZ optical depth degeneracy and future constraints on local primordial non-Gaussianity}

\newcommand{\uiuc}{Illinois Center for Advanced Studies of the Universe \& Department of Physics, University of Illinois at Urbana-Champaign, Urbana, IL 61801, USA}

\newcommand{\OSUCCAP}{Center for Cosmology and AstroParticle Physics (CCAPP), The Ohio State University, 191 West Woodruff Ave, Columbus, OH 43210, USA}

\author{Avery~J.~Tishue}
\email{atishue@illinois.edu}
\affiliation{\uiuc}

\author{Charuhas Shiveshwarkar}
\affiliation{\OSUCCAP}

\author{Gilbert Holder}
\affiliation{\uiuc}

\date{\today}

\begin{abstract}
Recent reconstructions of the large-scale cosmological velocity field with kinetic Sunyaev Zeldovich (kSZ) tomography have returned an amplitude that is low with respect to the halo model prediction, captured by the kSZ velocity reconstruction bias $b_v <1$. This suggests that common choices for modeling the galaxy-electron cross correlation have systematically overestimated the true power, at least over scales and redshifts used in the velocity reconstruction measurements. In this paper, we study the implications of this overestimation for constraints on local-type primordial non-Gaussianity in current and near-future cosmological surveys. For concreteness, we focus on kSZ velocity reconstruction from a Vera Rubin Observatory-like survey in tandem with contemporary cosmic microwave background measurements. Assuming standard choices for the fiducial model of the small-scale galaxy-electron cross correlation, we find that upcoming kSZ tomography measurements can significantly improve constraints on local primordial non-Gaussianity via measurement of scale-dependent galaxy bias, in broad concordance with previous studies of the application of kSZ tomography to primordial non-Gaussianity. However, when we instead modify the assumed galaxy-electron cross-spectrum to be consistent with recent measurements of the velocity reconstruction bias, this picture can change appreciably. 
Specifically, we find that if the inferred suppression of galaxy-electron power persists at higher redshifts $z\gtrsim 1$, kSZ-driven improvement in local primordial non-Gaussianity constraints may be less significant than previously estimated. We explore how these conclusions depend on various modeling and experimental assumptions and discuss implications for the emerging program of kSZ velocity reconstruction.

\end{abstract}

\maketitle

\section{Introduction\label{sec:intro}}

The hunt for insights into the primordial universe is a cornerstone of modern cosmology. The leading paradigm for generating appropriate initial conditions for the hot big bang cosmology, an inflationary epoch~\cite{Guth:1980zm,Linde:1981mu,Albrecht:1982wi,Mukhanov:1981xt,Hawking:1982cz}, has been enormously successful. Inflation drives the universe toward flatness, brings our observable patch into past-causal contact, and provides an elegant mechanism by which to generate a nearly scale-invariant spectrum of near-gaussian curvature perturbations in the early universe with which to seed the large-scale structure we observe today~\cite{Planck:2018jri, Linde:1981mu}. 
Despite this success, a precise microphysical picture of the early universe, inflation or otherwise, remains elusive. Indeed, seeking signals from a putative inflationary epoch, and the primordial universe more broadly, is a key science driver for many concerted experimental efforts presently underway in cosmology: Cosmic Microwave Background (CMB) experiments (e.g. the Simons Observatory~\cite{SimonsObservatory:2018koc}) and large-scale structure (LSS) surveys (such as the Vera Rubin Observatory (LSST)~\cite{LSSTScience:2009jmu} and SPHEREx~\cite{SPHEREx:2014bgr}) as well as gravitational wave observatories (e.g.~\cite{LISA:2017pwj}),  alike, are designed, in part, to search for clear signatures from the primordial universe. In parallel, a broad theoretical program has been developed to understand unique signatures and investigate optimal probes of new physics from the early Universe (see e.g.~\cite{Achucarro:2022qrl} for a review).

Among the most direct and theoretically compelling of these probes is primordial non-Gaussianity (PnG)~\cite{Bartolo:2004if, Komatsu:2009kd, Alvarez:2014vva}. The simplest realizations of inflation utilizing a single field generically predict near-trivial deviations from Gaussianity \cite{Achucarro:2022qrl} -- in other words, the primordial curvature perturbation in the simplest single-field inflation models is a Gaussian random variable at all scales. On the other hand, other models of inflation, for example multifield models~\cite{Chen:2009we, Chen:2009zp}, curvaton scenarios~\cite{Linde:1996gt}, non-Bunch-Davies vacuum states~\cite{Holman:2007na}, deviations from slow-roll \cite{Chen:2006xjb, Chen:2008wn, Flauger:2010ja, Adshead:2011bw, Adshead:2011jq, Miranda:2012rm, Adshead:2012xz, Adshead:2013zfa}, etc., can lead to non-Gaussianities of different amplitudes and scale-dependencies~\cite{Achucarro:2022qrl}. The type and extent of primordial non-Gaussianity are model-dependent and serve as distinguishing features of different inflationary models. In this way, PnG can be used as a theoretical testbed, providing a window into the detailed physics of the inflationary epoch and the primordial Universe more broadly.

In this work, we focus on \textit{local} primordial non-Gaussianity (LPnG)~\cite{Komatsu:2001rj}, which is typically characterized by a primordial curvature $\zeta$ which can be expressed as a local expansion in a Gaussian random field in real space~\cite{Lyth:2001nq, Sasaki:2006kq, Dvali:2003em}:
\begin{eqnarray}
\zeta(\textbf{x}) &=& \zeta_{g}(\textbf{x}) + \frac{3}{5}\fnlloc\left(\zeta_{g}(\textbf{x})^2 - \langle\zeta_{g}^2\rangle\right) \label{eq:LPnG_def}\ ,
\end{eqnarray}
where $\zeta_{g}$ is a Gaussian random field and we have dropped high order terms in the expansion (e.g. $g_{NL}$) for simplicity. The coefficient $\fnlloc$ is model-dependent and parametrizes the amplitude of the primordial bispectrum. LPnG is important from a theoretical point of view because it is a distinguishing feature of inflationary models with at least one additional light field (i.e. with mass $m \ll H$, where $H$ is the inflationary Hubble rate) beyond the inflaton~\cite{Achucarro:2022qrl,Meerburg:2019qqi,Tanaka:2011aj,Creminelli:2004yq}. A clear detection of non-zero LPnG from galaxy clustering and/or CMB anisotropies would convincingly rule out all single-field inflationary models and thus be a smoking gun for multi-field inflation~\cite{Pajer:2013ana, dePutter:2015vga, Babich:2004gb, Creminelli:2004yq}. 

Recently, reconstruction of the long-wavelength cosmological peculiar velocity field using kinetic Sunyaev Zeldovich (kSZ) tomography has emerged as a promising technique to augment probes of primordial non-Gaussianity \cite{Smith:2018bpn,Munchmeyer:2018eey,AnilKumar:2022flx, Adshead:2024paa,Krywonos:2024mpb,Lague:2024czc, Hotinli:2025tul, Lai:2025qdw}. The kSZ effect~\cite{1980ARA&A..18..537S,1980MNRAS.190..413S,1970CoASP...2...66S} is the scattering of CMB photons off high-energy electrons caught up in bulk flows. This scattering induces a secondary temperature anisotropy in the CMB and allows the reconstruction of the bulk velocity via the information encoded in the correlation between large-scale structure and the CMB anisotropies (see e.g. \cite{Smith:2018bpn,Bloch:2024kpn,McCarthy:2024nik}). The reconstructed velocity provides an additional tracer of matter density fluctuations, which can be combined with galaxy clustering measurements to mitigate cosmic variance on large scales~\cite{Seljak:2008xr} and seek the signal of PnG from scale-dependent galaxy bias~\cite{Dalal:2007cu,Smith:2010gx, Matarrese:2008nc, Smith:2011ub}. Fisher forecasts focusing on Stage III/IV surveys \cite{Munchmeyer:2018eey, Adshead:2024paa} have shown that this sample-variance cancellation can greatly improve constraints on $\fnlloc$ compared to galaxy survey data in isolation. Recent kSZ velocity reconstructions (see e.g. \cite{Krywonos:2024mpb,Lague:2024czc, Hotinli:2025tul, Lai:2025qdw, Bloch:2024kpn,McCarthy:2024nik,Kvasiuk:2025mtw}) have seen rapid improvements in constraining power: for example,~\citet{Lague:2024czc} used data from Planck~\cite{Planck:2018nkj} and the Atacama Cosmology Telescope Data Release 5 (ACT DR5) \cite{Naess:2020wgi} in tandem with the Sloan Digital Sky Survey Baryon Oscillation Spectroscopic Survey DR12 \cite{BOSS:2012dmf, BOSS:2016wmc}, detecting the galaxy-velocity cross correlation at $\sim\!7\sigma$ and constraining local primordial non-Gaussianity $\fnlloc = -90^{+210}_{-350}$ ($68\%$ CL), and more recent reconstructions \cite{Hotinli:2025tul,Lai:2025qdw} using ACT DR5 and the Dark Energy Spectroscopic Instrument Legacy Imaging Survey (DESI-LS)~\cite{DESI:2018ymu} made $>\!11\sigma$ measurements of the galaxy-velocity correlation, with Ref.~\cite{Hotinli:2025tul} reaching $\fnlloc = -39^{+40}_{-33}$.

However, a well-known technical challenge arises when attempting to use kSZ tomography for cosmological inference: the kSZ signal depends on the small-scale ($k_S \gtrsim 1 \,\rm{Mpc}^{-1}$) details of the galaxy-electron spectrum, which must be modeled in the absence of some independent measurement.\footnote{See e.g. Refs.~\cite{Madhavacheril:2019buy, AnilKumar:2025wyt} for proposed methods to do this based on fast radio burst and higher point correlation functions respectively.} This, in turn, manifests as an amplitude degeneracy on the reconstructed large-scale velocity field \cite{Smith:2018bpn}. This is known as the kSZ optical depth degeneracy~\cite{Battaglia:2016xbi,Flender:2016cjy,Louis:2017hoh,Soergel:2017ahb}, and in the context of cosmological applications it captures the uncertainty in $P_{ge}(k_S)$, the small-scale galaxy-electron cross spectrum, which is influenced by complicated astrophysical processes such as baryonic feedback~\cite{Dave:2000vh,Fukugita:2004ee,Cen:2006by,Hadzhiyska:2024qsl,Hadzhiyska:2025mvt,Bigwood:2025ism,Siegel:2025frt,Bigwood:2025kur,ACT:2025llb}. This uncertainty is captured by a parameter $b_v$, called the kSZ velocity reconstruction bias~\cite{Smith:2018bpn}, which relates the amplitude of the velocity field reconstructed with kSZ, $\hat{v}$, to the true underlying velocity field, $v$, via $\hat{v} \propto b_v v$. Moreover, in forecasts, overestimating $P_{ge}$ in the velocity reconstruction also overestimates the projected kSZ signal-to-noise ratio, thereby also overestimating the consequent improvement in constraining power.

Interestingly, recent analyses \cite{Hotinli:2025tul,Lai:2025qdw} have consistently found a $b_v$ well below unity at high significance, yielding $b_v \approx 0.4$. This indicates that the fiducial $P_{ge}$ models adopted in the velocity reconstructions have systemically overestimated the true galaxy-electron cross-correlation. This discrepancy highlights an important possibility: if standard modeling choices for $P_{ge}$ have been overestimates, then so have forecasts of the velocity reconstruction signal-to-noise, and projected constraints $\sigma(\fnlloc)$ have been optimistic. To what extent, however, depends crucially on the behavior of $P_{ge}$ as a function of scale and cosmic history. The kSZ analyses in Refs.~\cite{Hotinli:2025tul,Lai:2025qdw} have reconstructed the velocity field for $z\lesssim 1$, but the value of $b_v$ at higher redshifts has not yet been inferred.

In this paper, we explore the possible implications of recent low $b_v$ measurements for future efforts to constrain $\fnlloc$ with kSZ tomography. The key ingredient in our analysis is the range of assumptions made about the suppression of $P_{ge}$ compared to common fiducial choices. To distill our analysis, we focus on simple phenomenological models of $P_{ge}$ that consistently reproduce the inferred suppression of small-scale galaxy-electron power \cite{Hotinli:2025tul,Lai:2025qdw}. We then evaluate how such modifications affect the velocity reconstruction noise, and in turn, the forecasted constraints on $\fnlloc$. In our analysis, we consider a range of experimental settings and modeling choices, and we consistently find that the most salient input is the amplitude of the small-scale galaxy-electron correlation at high redshifts, $z\gtrsim 1$, where access to larger scale modes improves the efficacy of sample variance cancellation. On one hand, if the overestimation of $P_{ge}$ is isolated to low redshifts ( $z\lesssim 1$, where recent measurements of $b_v<1$ \cite{Hotinli:2025tul,Lai:2025qdw} have been made), such that at higher redshifts the amplitude of $P_{ge}$ is well-described by commonly used models, then future reductions in $\sigma(\fnlloc)$ from kSZ tomography are barely affected. On the other hand, if the suppression of $P_{ge}$ is also persistent at higher redshifts, then we find that  future improvements in $\sigma(\fnlloc)$ from kSZ tomography may be significantly less than previously understood.

The remainder of this paper is organized as follows. In Sec.~\ref{sec:fNL}, we give a brief overview of local primordial non-Gaussianity and observational imprint on the large-scale galaxy bias. We outline the essential details of galaxy and kSZ tomography, including the kSZ optical depth degeneracy in Sec.~\ref{sec:gal_ksz_tomography}. In Sec.~\ref{sec:forecasts}, we describe our forecast methodology and analysis choices, including models of the galaxy-electron power spectrum and assumptions about cosmological surveys. We present our results in Sec.~\ref{sec:results}, and discuss our findings in Sec.~\ref{sec:discussion}.

\section{Local Primordial non-Gaussianity \label{sec:fNL}}
Our focus is on local-type primordial non-Gaussianity, in which the primordial curvature fluctuation $\zeta$ is parameterized in real space by a gaussian component $\zeta_g$ and a non-Gaussian deviation whose amplitude is controlled by the parameter $\fnlloc$~\cite{Achucarro:2022qrl, Babich:2004gb}:
\begin{align}
\zeta(\mathbf{x}) = \zeta_g(\mathbf{x}) + \frac{3}{5}\fnlloc(\zeta_g(\mathbf{x})^2 - \langle \zeta_g(\mathbf{x})^2\rangle) \label{eq:fNLloc_defn}\ .
\end{align}
With local-type primordial non-gaussianity, the momentum space three-point function (the bispectrum) has an amplitude that peaks in the so-called squeezed configuration in which the three momenta have amplitudes $k_1 \ll k_2, k_3$. In this case, the bispectrum takes the form~\cite{Achucarro:2022qrl, Babich:2004gb}
\begin{align}
  \langle \zeta(k_1\rightarrow 0)\zeta(k_2) \zeta(k_3) \rangle \simeq \fnlloc P_{\zeta}(k_1)P_{\zeta}(k_{2}) \times \nonumber\\(2\pi)^{3}\delta^{(3)}(\mathbf{k}_1+ \mathbf{k}_2+\mathbf{k}_3) \label{eq:fNLbispectrum}\ .
\end{align}
This type of primordial non-Gaussianity is associated with a characteristic signature in the late universe in the form of a scale-dependent galaxy-bias with an amplitude that grows quadratically as the inverse-wavenumber goes to arbitrarily large distance scales~\cite{Dalal:2007cu, Slosar:2008hx, Matarrese:2008nc}:
\begin{equation}
    \Delta b(k,z) = \fnlloc \frac{b_{\phi}(z)}{2 k^2 T(k) D(z)/3H_0^2 \Omega_m}\ ,
    \label{eq:bNG_defn}
\end{equation}
where the transfer function $T(k)$ is normalized to unity as $k\rightarrow 0$ and the linear growth factor $D(z)$ is normalized such that $D(z) = 1/(1+z)$ in the matter-dominated era. Due to the $1/k^2$ scaling in Eq.~\eqref{eq:bNG_defn}, the scale-dependent galaxy bias signal peaks at large, near-horizon scales where the contribution of gravitational non-linearities to galaxy clustering is sub-dominant. The presence of such a large observable signal at linear/mildly non-linear scales makes LPnG especially amenable to detection by current and upcoming large-scale galaxy surveys.  

The factor $b_{\phi}(z)$ in Eq.~\eqref{eq:bNG_defn} denotes the response of galaxy abundance $n_g$ to the variance of the smoothed matter-density field $\sigma_8$~\cite{Dalal:2007cu, Slosar:2008hx, Matarrese:2008nc},
\begin{eqnarray*}
b_{\phi}(z) = 2\frac{\partial\log n_{g}}{\partial\log\sigma_{8}}\ ,
\end{eqnarray*}
and is sensitive to the complicated, small-scale physics of galaxy formation which may not be expressible analytically. Consequently, one needs to marginalise over $b_{\phi}$ in order to obtain meaningful constraints on $\fnlloc$ from galaxy clustering observations. In the particular case where the tracer abundance $n_{g}$ is a function only of peak height $\delta_{c}/\sigma_{8}$, with $\delta_{c}=1.686$ being the threshold overdensity of spherical collapse in an Einstein-deSitter universe, one can show that $b_{\phi}$ is related to the galaxy bias $b_{g}$ as~\cite{Dalal:2007cu} 
\begin{eqnarray}
    \label{eq:universality}
    b_{\phi} = 2(b_{g}-1)\delta_{c}\ .
\end{eqnarray} 
The relation~\eqref{eq:universality} is known as the \textit{universal mass function ansatz}~\cite{Dalal:2007cu, Slosar:2008hx}. One the other hand, most galaxies and even dark-matter halos selected by properties beyond their mass in general do not obey the universal mass function ansatz~\cite{Desjacques:2016bnm, Barreira:2022sey, Barreira:2020kvh, Lazeyras:2022koc} and show deviations from Eq.~\eqref{eq:universality} which arise not just from peculiarities of the small-scale physics of galaxy/halo formation but also from non-trivial features of the tracer selection function~\cite{Dalal:2025eve,Sullivan:2025fie,Shiveshwarkar:2025nac}. In this work, for simplicity and to compare with previous results \cite{Hotinli:2025tul,Lai:2025qdw}, we assume the universal mass function ansatz (Eq.~\eqref{eq:universality}).

\section{Galaxy and kSZ tomography \label{sec:gal_ksz_tomography} } 
\subsection{Observables \label{subsec:observables}}
In cosmological applications of kSZ tomography, the goal is to extract cosmological information from the large-scale velocity field by leveraging the correlations induced in the CMB and large-scale structure due to kSZ scattering. To do so, one can  write a quadratic estimator for the radial peculiar velocity mode $\hat{v}_r$ in terms of the observed CMB temperature $T$ and galaxy overdensity $\delta_g$, $\hat{v}_r \sim  \delta_g T_{\mathrm{kSZ}} $. Working in the simplified box geometry and long wavelength velocity reconstruction formalism outlined in Ref.~\cite{Smith:2018bpn}, the universe at redshift $z$ is modeled as a periodic 3D box of comoving volume $V$, so that the velocity estimator is
\begin{align}
    \hat{v}_r(k_L) = N_{\hat{v}_r}(k_L)\frac{K(z)}{\chi(z)} &\int \frac{d\mathbf{k}_S}{(2\pi)^3} \frac{d \boldsymbol{\ell}}{(2\pi)^2}\, \delta_g^*( \boldsymbol{k}_S) T^*( \boldsymbol{\ell})  \label{eq:v_estimator}\\
    & \times \frac{P_{ge}(k_S)}{P_{gg}^{\mathrm{obs}}(k_S, k_{L,r}) C_{\ell=k_s \chi(z)}^{TT, \mathrm{obs}}}\nonumber\\
    &\times(2\pi)^3 \delta^3(\boldsymbol{k}_S + \boldsymbol{\ell}/\chi(z) + \boldsymbol{k}_L)\ . \nonumber
\end{align}
Here $\chi(z)$ is the comoving distance out to redshift $z$ and the prefactor $K(z)$ depends on the Thompson cross section $\sigma_T$, the electron number density $n_{e,0}$, free electron fraction $x_e$, and optical depth $\tau$ via $K(z) = \sigma_T n_{e,0} x_{e}(z) e^{-\tau(z)} (1+z)^2$. The integral is over small scales $k_S \sim 1/\rm{Mpc}$ and high multipole $\ell \sim 4000$, and the velocity field is reconstructed on large scales $k_L$; we outline the distinction between $k_S$ and $k_L$ further in Sec.~\ref{sec:forecasts}.  $P_{ge}$ is the galaxy-electron cross-spectrum, and  $P_{gg}^\mathrm{obs}(k_S, k_{L,r})$ and  $C_{\ell}^{TT,\mathrm{obs}}$ are the observed galaxy and CMB temperature power spectra respectively. The weight of the estimator, $P_{ge}/P_{gg}^{\rm{obs}} C_{\ell}^{\rm{obs}}$, is chosen so that $\hat{v}_r$ is a minimum variance estimator of the true underlying radial velocity field, $v_r$. 

The observed galaxy power spectrum can be anisotropic due to photometric redshift errors degrading the measurement along the radial direction, so that $P_{gg}^{\rm{obs}}$ depends explicitly on the radial component of the large-scale mode to be reconstructed, $k_{L,r} = \mu k_L$ where $\mu = \hat{k}_L\cdot \hat{r}$. Then, our model for the observed galaxy power spectrum is
\begin{align}
    P_{gg}^\mathrm{obs}(\mu,k,z ) = W^2(\mu,k,z ) b_g^2(\mu,k,z ) P_{mm}(k,z ) + N_g(z)\ ,
\end{align}
where $P_{mm}$ is the matter power spectrum, $b_g$ is the galaxy bias, the galaxy shot noise is $N_g = 1/n_g$ for galaxy number density $n_g$, and the photo-z function is $W = \exp (-k_{L,r}^2 \sigma_z^2 /2H^2(z))$, where  $H(z)$ is the Hubble rate and $\sigma(z) = \bar{\sigma}_z (1+z)$ is the photometric redshift error characterized by redshift scatter parameter $\bar{\sigma}_z$. Finally, the noise on the reconstructed velocity field, $N_{\hat{v}_{r}}(k_L)$, is  
\begin{eqnarray}
  N_{\hat{v}_{r}}(k_L,z) &=&  \frac{\chi^2(z)}{K^2(z)}\times\nonumber\\
  &&\left[\int \frac{dk_S}{2\pi} \, k_S \frac{P^2_{ge}(k_S,z)}{P_{gg}^{\mathrm{obs}}(k_S, k_{L,r},z )C_{\ell=k_s \chi(z)}^{TT, \mathrm{obs}}} \right]^{-1}\ . \label{eq:ksz_noise}
\end{eqnarray}

Reconstruction of the (radial) velocity field $v_r$ with kSZ tomography provides an additional tracer of the matter density fluctuation $\delta_m$, where the two are related in linear theory via
\begin{align}
    v_r = \mu (faH/k)\delta_m\ ,
\end{align}
where $f(z) = -d\log D(z)/d\log (1+z)$ is the cosmic growth rate. The search for the scale-dependent galaxy bias signal is challenged by the fact that it is greatest on large cosmological scales ($k \rightarrow 0$), where cosmic variance can limit the precision achievable in constraining galaxy bias. Combining the galaxy and velocity measurements, then, facilitates cosmic variance cancellation \cite{Seljak:2008xr} for measuring the scale-dependent galaxy bias signal, $\Delta b_g(k) \propto (P_{gg}/P_{gv_r})$. Together, the galaxy and velocity power and cross-spectra are
\begin{align}
    &P_{gg}(\mu, k,z) = W^2 (\mu,k,z )b_g^2(\mu,k,z) P_{mm}(k,z) + N_g(z)\ , \\
    &P_{gv_r}(\mu, k,z) =  W(\mu,k,z ) b_g(\mu,k,z ) b_v(z) \mu \frac{faH}{k}P_{mm}(k,z)\ , \\
    &P_{v_r v_r}(\mu,k,z) = b_v^2(z) \mu^2 \left(\frac{faH}{k} \right) ^2 P_{mm}(k,z) + N_{\hat{v}_r}(\mu,k,z)\ .
\end{align}
Here, the factor $b_v(z)$ is the `kSZ velocity bias', which arises from ignorance of the true, small-scale galaxy-electron power spectrum, $P_{ge}(k_S)$. This is a key point in this paper, which we now turn to in more detail.

\subsection{kSZ optical depth degeneracy \label{subsec:opticaldepth} } 
As  shown in Eq.~(\ref{eq:v_estimator}), the reconstruction of the velocity field can be expressed in the quadratic estimator formalism as an integral over the weighted small-scale galaxy and CMB maps, with the respective weights being proportional to the small-scale galaxy-electron cross-spectrum, $P_{ge}(k_S)$. In Ref.~\cite{Smith:2018bpn}, it was very elegantly shown that kSZ tomography actually measures a squeezed bispectrum of the form $\langle \delta_g \delta_g T \rangle$, where $\delta_g$ and $T$ are galaxy and CMB temperature modes respectively, and this bispectrum is proportional to two power spectra: $\langle \delta_g \delta_gT \rangle \propto P_{ge}(k_S) P_{gv}(k_L)$. From the point of view of cosmology, one goal of kSZ tomography is to extract useful information from an independent probe of the large-scale velocity field, $v_{k_L}$, here appearing in the form of the galaxy-velocity cross spectrum $P_{g\hat{v}}$. However, from the kSZ measurement of the squeezed bispectrum, we are faced with a perfect amplitude degeneracy between our quantity of interest, $P_{gv}(k_L)$, and the small-scale galaxy-electron cross-spectrum $P_{ge}(k_S)$. This is known as the kSZ optical depth degeneracy~\cite{Smith:2018bpn}. 

In this context, in the absence of an independent measurement of free electrons on small scales, the kSZ velocity reconstruction requires as input some  fiducial model, $P^{\rm{fid}}_{ge}(k_S)$. If this model differs from the true galaxy-electron cross correlation, $P_{ge}^{\rm{true}}$, the kSZ-reconstructed velocity $\hat{v}$ is biased compared to the true velocity field, $\hat{v} = b_v v^{\rm{true}}$. This kSZ velocity construction bias, $b_v$, is determined by the mismatch between the fiducial galaxy-electron power spectrum and the true galaxy-electron power spectrum on small scales $k_S$, taking the form
\begin{align}
\label{eq:bv}
b_{v} &= \frac{\int dk_S F(k_S) P^{\mathrm{true}}_{ge}(k_S)}{\int dk_S  F(k_S) P^{\mathrm{fid}}_{ge}(k_S)}\ , 
\end{align}
where 
\begin{align}
F(k_S) &= k_S \frac{P^{\mathrm{fid}}_{g e}(k_S)}{P^{\mathrm{obs}}_{g g}(k_S) C_{\ell = k_S \chi}^{TT, \mathrm{obs}} }\ ,  \label{eq:F(k)}
\end{align}
such that $b_v =1$ when the model for the galaxy-electron cross correlation is equal to the true one, at least over the range of scales $k_S$ used in the velocity reconstruction. In cosmological analyses, one must marginalize over $b_v$ in the same way that one marginalizes over galaxy biases. Note, however, that $b_{v}$ captures an amplitude uncertainty in $P_{gv}$ that is inherited from ignorance of the small-scale galaxy-electron power spectrum. In that sense, the reconstruction bias $b_{v}$ is conceptually different from the galaxy bias $b_{g}$, even though both manifest as a multiplicative bias in modeling power spectra. In particular the reconstruction bias $b_v$ is scale-independent, and does not get a scale-dependent contribution from $\fnlloc$.

\section{Forecasts \label{sec:forecasts}}  
In this section, we lay out our forecast methodology, including the range of assumptions we make about the galaxy-electron spectra as well as the details of the cosmological surveys we consider.

\subsection{Galaxy-Electron Power Spectra \label{subsec:Pge}}

As discussed in the previous section, the fact that $P_{ge}(k_S)$ is not known a priori formally leads to the reconstructed velocity being biased by $b_v$. However, more broadly,  it means that the reconstruction noise $N_{\hat{v}}$ has a `theory' error bar, which in turn can significantly affect forecasted parameter constraints.

To this point, crucially, recent analyses of kSZ velocity reconstruction with ACT and DESI data have \textit{not} found $b_v \approx 1$, and instead have independently returned values of $b_v<1$ at high significance:  Ref~\cite{Hotinli:2025tul} finds $b_v = 0.45^{+0.06}_{-0.05}$, while Ref.~\cite{Lai:2025qdw} finds a very similar result, $b_v = 0.39 \pm 0.04$, despite differences in analysis choices and techniques. That is, $b_v$ is found to be consistently well below unity at high significance. While here we have quoted the results when the redshift range $0.4 < z \lesssim 1$ is collapsed into a single bin \cite{Hotinli:2025tul, Lai:2025qdw},~\citet{Lai:2025qdw} also showed that a low $b_v$ of roughly the same magnitude is consistent across the entire redshift range $0.4 < z \lesssim 1$ when binned more finely ($\Delta z = 0.035$). 

Both analyses \cite{Hotinli:2025tul, Lai:2025qdw} adopt the Battaglia profile~\cite{Battaglia:2016xbi} (also often called the AGN model) for $P_{ge}$, which is a standard choice in studies of kSZ tomography~\cite{Smith:2018bpn}; other choices are possible, although among several common choices, such as the shock heating (``SH'') \cite{Battaglia:2016xbi} and Universal profiles~\cite{Komatsu:2001dn}, the AGN estimate for $P_{ge}$ is lower \cite{Smith:2018bpn} due to stronger feedback \cite{Battaglia:2016xbi}. Because $b_v$ is an integrated quantity, it is only clear from these measurements that the fiducial models for $P_{ge}$ on small scales have been overestimations, while the exact form of this overestimation as a function of scale is not clear.

In this paper, our primary goal is to understand what such overestimation could reasonably imply about future prospects for probing $\fnlloc$ with kSZ tomography. Our focus here is not the precise details of the form of $P_{ge}^{\rm{true}}(k_S)$. Rather, our aim is to characterize, in a broad sense, how the suppression of $P_{ge}$ vis-à-vis the AGN model, as implied by Refs.~\cite{Hotinli:2025tul,Lai:2025qdw}, impacts the projected constraining power of forthcoming measurements. For this reason, we will adopt a simple approach, in which we define a toy model in fourier space meant to capture, in the most general sense, the key feature, i.e. the suppression of power. Concretely, in this work, we consider two different models for this suppression of power, which for convenience and for comparison with Refs.~\cite{Hotinli:2025tul,Lai:2025qdw}, we define with respect to the AGN model: either a flat suppression, or an exponential suppression that decrements power on the smallest scales, 
\begin{align}
    P_{ge}^{\rm{(flat)}}(z,k_S) &= A^2(z) P_{ge}^{\rm{(AGN)}} (z,k_S) \label{eq:Pge_suppress_models_flat}\ , \\
    P_{ge}^{(k_*)}(z,k_S) &= e^{-k_s/k_*(z)} P_{ge}^{\rm({AGN)}} (z,k_S) \label{eq:Pge_suppress_models_roll}\ .
\end{align}
Here, the quantity $A\leq 1$ qualitatively captures some scale-free overestimation of power. A slightly more general approach is to relax the simplifying assumption of scale-free suppression, and instead posit that the galaxy-electron power is suppressed below some scale, which is here captured by the comoving wavenumber $k_*$ in Eq.~(\ref{eq:Pge_suppress_models_roll}). These toy models are useful, allowing simple comparison with common fiducial models, but we emphasize that they are not meant to correspond to precise realizations of complex astrophysical processes. There are many complex inputs in computing $P_{ge}(k_S)$, such as baryonic feedback, small-scale galaxy bias, non-linear clustering, etc., and hence there are many ways to imagine modeling (or mis-modeling) $P_{ge}(k_S)$. Choices more detailed than Eqs.~(\ref{eq:Pge_suppress_models_flat},\ref{eq:Pge_suppress_models_roll}) are possible; this approach, however, facilitates the exploration of suppressed galaxy-electron power in a simple, tuneable, phenomenological prescription.

In this prescription, we will consider the following four cases, two for each model in Eqs.~(\ref{eq:Pge_suppress_models_flat}) and (\ref{eq:Pge_suppress_models_roll}):
\begin{itemize}
    \item flat suppression, only over the redshift range, $z<1$
    \item flat suppression, over the entire redshift range in our analysis
    \item exponential suppression, with a constant comoving suppression scale $k_*(z)=k_*$ over all redshifts
    \item exponential suppression, with a redshift-dependent comoving suppression scale, $k_*(z)\propto 1/(1+z)$.
\end{itemize}
The first case, a flat suppression only at low redshifts, is achieved in practice by setting $A^2(z<1) = 0.45$, to match Refs.~\cite{Hotinli:2025tul}, and $A(z>1)=1$, recovering the AGN model at high redshifts. While a sudden step in $P_{ge}$ near $z=1$ is unphysical, this is the simplest, minimal choice that imposes consistency with Refs.~\cite{Hotinli:2025tul,Lai:2025qdw}, while making no extrapolations to higher redshifts. The second case instead supposes that the measured suppression continues out to higher redshifts, still in a scale-independent way, i.e. we take $A^2(z)=0.45$ for all $z$. Together, these two choices serve as a simple diagnostic of the relative impact that high- vs. low-redshift suppression of $P_{ge}$ (and the corresponding increase in $N_{\hat{v}_r}$) has on $\fnlloc$ constraints, while drawing contact with recent measurements. 

In the third and fourth cases, the exponential suppression model explores the effect of scale-dependence in the suppression of $P_{ge}$, which we take to vanish on large scales where it is assumed electrons trace the total matter field \cite{Smith:2018bpn}. In the model with a constant comoving suppression scale, $k_* = \rm{const}$, we chose a value of $k_* = 2.38/\rm{Mpc}$ which reproduces $b_v = 0.45$ at the central redshift $z_* = 0.7$ found by Refs.~\cite{Hotinli:2025tul,Lai:2025qdw}, when assuming galaxy and CMB survey details following their analysis.\footnote{We use a single redshift bin from $z=0.4$ to $z=1$, centered at $z_*=0.7$, with a galaxy number density $3.2\times10^{-4} \, [\mathrm{Mpc}]^{-3}$, photo-z scatter $\bar{\sigma}_z = 0.03$, and a CMB resolution and noise of $1.4'$ and $7 \mu K'$ respectively.} While this case assumes a redshift-independent suppression of $P_{ge}$, the response of the noise $N_{\hat{v}_r}$ is redshift dependent due to the momentum conserving delta function imposing $k_S = \ell/\chi(z)$, which causes the range of scales $k_S$ that the reconstruction noise is sensitive to depend on redshift. Finally, we consider a case in which the comoving suppression scale varies over cosmic history, $k_*(z)$. In particular, we consider the possibility of some characteristic, fixed \textit{physical} scale in the system, i.e. a wavenumber $k_{*,\rm{phys}}$, below which galaxies and electrons are decorrelated, which in turn yields a comoving suppression scale that goes as $k_*(z) \propto 1/(1+z)$. To again ensure consistency with recent measurements, we choose the proportionality constant to reproduce the measurement of \cite{Hotinli:2025tul}, i.e. we use $k_*(z) = k_{z_*}(1+z_*)/(1+z)$ with $k_{z_*} = k_* = 2.38/\rm{Mpc}$ appropriately chosen to yield $b_v(z_*=0.7) = 0.45$. 

\begin{figure}[t!]
    \centering
    \includegraphics[width=1.\linewidth]{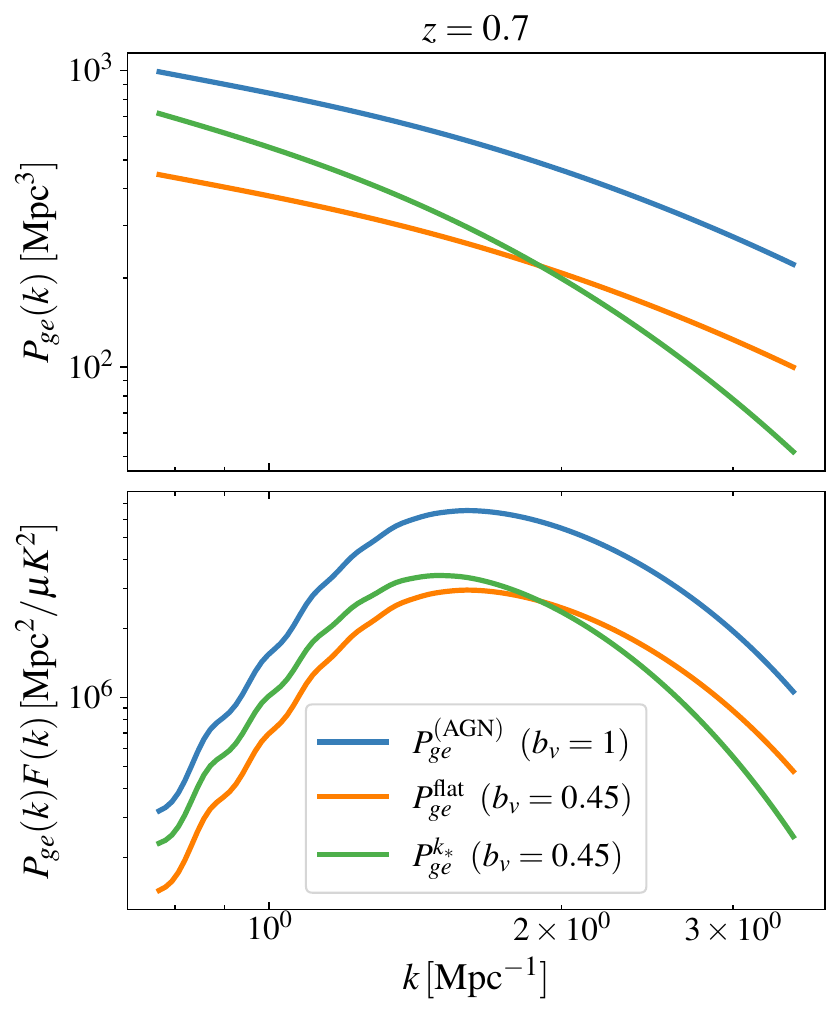}
    \caption{ Comparing galaxy-electron power spectra models (upper panel) and their corresponding effect on the $b_v$ integrand (lower panel) at $z=0.7$. The AGN model (blue line) is compared to a flat suppression (orange line), Eq.~(\ref{eq:Pge_suppress_models_flat}), and an exponential suppression (green line), Eq.~(\ref{eq:Pge_suppress_models_roll}). As detailed in the main body of the text, the model parameters of the suppressed orange and green curves, $A^2=0.45$ and $k_* = 2.38 / \rm{Mpc}$ respectively, are chosen to reproduce the value of $b_v=0.45$ of Ref.~\cite{Hotinli:2025tul} (with respect to the AGN model, hence the blue curve has $b_v=1$ by definition), when approximating their CMB and galaxy survey details. Note the upper and lower panels share a legend and a horizontal axis. }
    \label{fig:pge_and_bv_integrand}
\end{figure}

In Fig.~\ref{fig:pge_and_bv_integrand}, we compare these suppressed $P_{ge}$ models (orange and green) to the AGN model (blue), and we also show the corresponding effect on the integrand of $b_v$, Eq.~(\ref{eq:bv}) (which is essentially the same as the reconstruction noise integrand, Eq.~(\ref{eq:ksz_noise})). We show the comparison in a bin centered at $z=0.7$, where the models are calibrated to yield $b_v=0.45$ (and for which the $k_*$ and $k_*(z)$ models are the same). Compared to flat suppression, the exponential suppression scenario comparatively decreases power more on at higher $k$, and less at lower $k$; correspondingly, the integrands controlling $b_v$ differ, as shown in the lower panel, but the main point is that both yield the same $b_v$ value (i.e., they give the same integrated suppression) compared to the AGN model.

\begin{figure*}
    \centering
    \includegraphics[width=1.\linewidth]{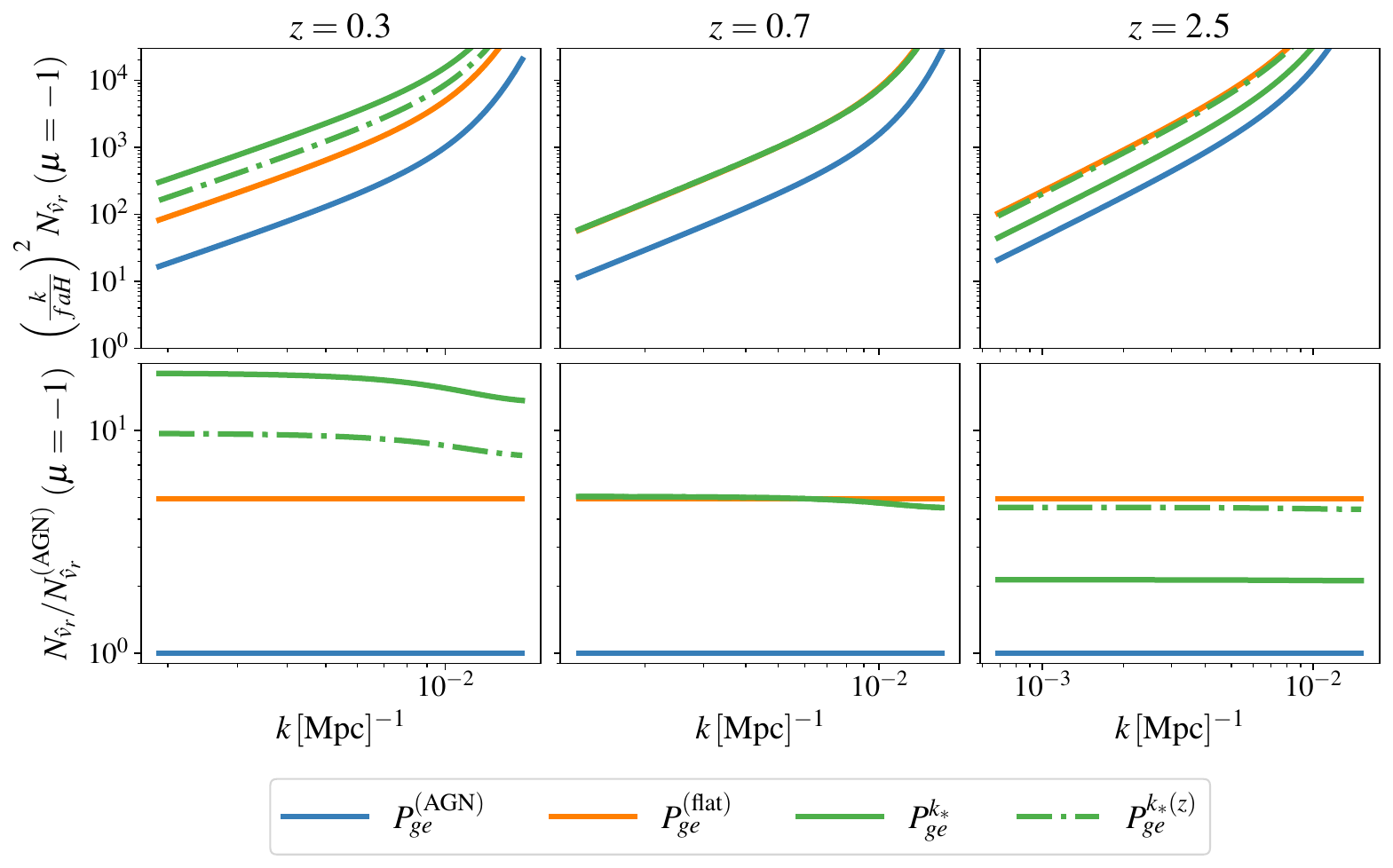}
    \caption{ The kSZ velocity reconstruction noise (upper panels), modulated by a factor $(k/faH)^2$ to correspond to the reconstruction noise on the density modes \cite{Smith:2018bpn,Munchmeyer:2018eey}, for varying $P_{ge}$ models, and the ratio of these reconstruction noises to that obtained from the AGN model (lower panel), for three redshifts $z\in[0.3,0.7,2.5]$ (left, center, and right panels respectively), for purely radial modes $\mu=-1$. The flat suppression model (orange, Eq.~(\ref{eq:Pge_suppress_models_flat})) and exponential suppression models (Eq.~(\ref{eq:Pge_suppress_models_roll}), green solid and dot-dashed) yield a larger reconstruction noise than the AGN model (blue). The model parameters $(A,k_*,z_*)$ are the same as in Fig.~\ref{fig:Nv_Pge_compare}, as described in the main body of the text. Note the solid and dot-dashed green curves lie exactly on top of one another in the center panel. Here, the CMB experiment is assumed to have a $1.4'$ resolution and noise level $\Delta_T = 4 \mu K'$, and the galaxy survey specifications are that of our LSSTY10 forcast, outlined further in Sec.~\ref{subsec:surveys} and Table~\ref{tab:LSST}.} 
    \label{fig:Nv_Pge_compare}
\end{figure*}

The effect of the different $P_{ge}$ scenarios on the velocity reconstruction noise is key to our analysis, which we show further in Fig.~\ref{fig:Nv_Pge_compare} (where we have multiplied the velocity noise by $(k/faH)^2$ to correspond to the reconstruction noise for the density modes~\cite{Smith:2018bpn,Munchmeyer:2018eey}). We examine several redshifts, $z \in [0.3,0.7,2.5]$, and show purely radial modes ($\mu =- 1)$ for simplicity. In the upper panels, we show the reconstruction noise for different $P_{ge}$ scenarios, while in the lower panel, we show the fractional increase in the noise due the suppression of $P_{ge}$ compared to the AGN model. The orange curves represent the flat suppression, with $A$ defined to reproduce $b_v =0.45$. The green curves show the exponential suppression models with a fixed comoving suppression scale $k_*$ (solid curves) and redshift dependent suppression scales $k_*(z)$ (dot-dashed curves). To understand these curves, it is useful to recall the scales to which the reconstruction noise is sensitive. The kSZ signal (i.e., the filter/weight functions in the $N_{\hat{v}}$ and $b_v$ integrands; see also Refs.~\cite{Smith:2018bpn,Hotinli:2025tul, Lai:2025qdw}) peaks around $\ell \sim 4000$, related to wavenumbers via $k_S = \ell/\chi(z)$ due to the momentum conserving delta function in the box geometry \cite{Smith:2018bpn}. Hence the $k_S$ integral bounds for the reconstruction noise moves to larger scales (smaller wavenumbers) at higher redshifts. At $z=0.7$, we have chosen the model parameters $A,k_*,z_*$ to yield $b_v(z=0.7)=0.45$. This is why the orange and green curves overlap in the central panel of Fig.~\ref{fig:Nv_Pge_compare}.\footnote{ Note, due to photometric redshift errors, the response of the reconstruction noise on large scales acquires a slight scale-dependence if the change to $P_{ge}(k_S)$ is scale-dependent, which can be seen from Eq.~(\ref{eq:ksz_noise}), and this is why there is a small step-like feature in the green curves in the lower panels, but not the orange curves.} However, at lower redshifts, e.g. $z=0.3$, the reconstruction noise is sensitive to smaller scales, that is, larger $k_S$,   where the exponential suppression factor ~$e^{-k/k_*}$ is greater. Hence, at lower redshifts, the exponential suppression (green, solid) is more severe than the flat suppression (orange), $P_{ge}^{k_*} < P_{ge}^{(\rm{flat})}$, and therefore the corresponding reconstruction noise is larger, $N_{\hat{v}_r}^{k_*}> / N_{\hat{v}_r}^{\rm{(flat)}}$. The trend reverses at high redshifts, where the window of scales for the reconstruction noise moves to smaller $k_S$ values, for which the exponential suppression is less severe. For the same reasons, the model with a redshift-dependent comoving suppression scale $k_*(z)\sim 1/(1+z)$ (i.e. constant $k_{*,\rm{phys}}$), in dot-dashed green, has a larger reconstruction noise at high redshifts than the model with a constant comoving suppression scale $k_* = \rm{const}$ (solid green). Comparing these two at redshifts $z>0.7$, as shown in the right panel of Fig.~\ref{fig:Nv_Pge_compare}, we have $k_*(z)<k_*$, and hence the exponential suppression is more extreme at higher redshifts, and therefore the reconstruction noise is larger. Conversely, for $z<0.7$, we have the opposite situation, $k_*(z)>k_*$, and the reconstruction noise for the redshift dependent model $k_*(z)$ is slightly smaller than that of the $k_* = \rm{const}$ model, as shown in the left panel of Fig.~\ref{fig:Nv_Pge_compare}.

Anticipating our main focus, the $\fnlloc$ forecasts that we present in Sec.~\ref{sec:results}, this behavior of the reconstruction noise in Fig.~\ref{fig:Nv_Pge_compare} is crucial. It will turn out that the high redshift information ($z>1$) is very important. From this, one can already deduce, for example, that the model with flat suppression at high redshifts will degrade the constraints the most compared to the AGN model, while the impact of the constant comoving suppression scale $k_*$ will be comparatively less significant.

\subsection{Survey Details \label{subsec:surveys}}
To focus our study, our primary analysis will consider the combination of LSST galaxy measurements and a $1.4'$ resolution CMB survey with a noise level that we vary in our forecasts. In our model for the measured CMB power spectrum, necessary for the computation of the kSZ velocity reconstruction noise, we follow the cross-ILC approach in \cite{Raghunathan:2023yfe}\footnote{ \url{https://github.com/sriniraghunathan/CMB_BAO_SNe_likelihoods/ }}, assuming the following contributions from foregrounds: the cosmic infrared background, the thermal SZ effect, the kSZ effect, weak lensing, radio sources, and dusty star forming galaxies. Additionally, the noise spectrum is defined as
\begin{align}
    N_{\ell} = \Delta_T \exp \left(\frac{\ell (\ell+1) \theta_{\mathrm{FWHM}}^2}{ 8 \ln(2)} \right)\ ,
\end{align}
for a noise amplitude $\Delta_T$ and CMB beam $\theta_{\mathrm{FWHM}}$. In our analysis, we will assume  a resolution $\theta_{\mathrm{FWHM}} = 1.4'$, and we will vary the noise level in the range $1<\Delta_T \, [\mu K' ]<12$. In computing the reconstruction noise, we integrate over the range of scales $k_S = \ell /\chi(z)$ by using $(\ell_{\rm{min}},\ell_{\rm{max}}) = (2000,9000)$, following Ref.~\cite{Hotinli:2025tul}. Modifying this range slightly does not change our conclusions. 
We compute the galaxy-electron spectrum and kSZ reconstruction noise using the public code  \texttt{hmvec}\footnote{\url{https://github.com/simonsobs/hmvec}} \cite{Smith:2018bpn}.

We assume the specifications of the LSSTY10 gold sample~\cite{LSSTScience:2009jmu}, with a redshift scatter of $\bar{\sigma}_z = 0.03$, divided into 8 redshift bins out to redshift $z = 2.8$. The galaxy number density and fiducial linear bias values are outlined in Table~\ref{tab:LSST}. Previous studies \cite{Munchmeyer:2018eey, AnilKumar:2022flx, Adshead:2024paa} have indicated that LSST is an excellent survey for the application of galaxy and kSZ tomography to primordial non-Gaussianity. The high number density of the LSST gold sample reduces the kSZ reconstruction and galaxy shot noise. The single-tracer constraining power of the gold sample already approaches the cosmic variance limit~\cite{Munchmeyer:2018eey}, whereas a multi-tracer approach with kSZ continues to benefit from higher number densities. Moreover, the relatively low photometric errors out to high redshift aids in the measurement of the small-scale modes used in the velocity reconstruction.

\begin{table}[b!]
\centering
\caption{Redshift bins $(z_{\mathrm{min}}, z_{\mathrm{max}})$, galaxy number densities $n_{g}$ [$(\mathrm{ Mpc})^{-3}$], linear galaxy biases $b^{\rm{lin}}$, and photo-z scatter parameter $\bar{\sigma}_z$ for the LSST forecasts in this work.}
\label{tab:LSST}
\begin{tabular}{c c|cc}
\hline
 &  & \multicolumn{2}{c}{LSST Gold Sample} \\
 &  & \multicolumn{2}{c}{$\bar{\sigma}_z = 0.03$} \\
$z_{\min}$ & $z_{\max}$ & $n_{g}\,[\mathrm{Mpc}]^{-3}$ & $b^{\rm{lin}}$ \\
\hline
0.0 & 0.2 & $6.94\times 10^{-2}$   & 1.00 \\
0.2 & 0.4 & $4.76\times 10^{-2}$   & 1.11 \\
0.4 & 0.6 & $3.14\times 10^{-2}$   & 1.24 \\
0.6 & 0.8 & $2.05\times 10^{-2}$   & 1.37 \\
0.8 & 1.0 & $1.33\times 10^{-2}$   & 1.50 \\
1.0 & 1.6 & $5.71\times 10^{-3}$   & 1.78 \\
1.6 & 2.2 & $1.46\times 10^{-3}$   & 2.21 \\
2.2 & 2.8 & $3.39\times 10^{-4}$   & 2.65 \\
\hline
\end{tabular}
\end{table}

As an alternative setup, we will also consider a scenario in which small-scale LSST measurements are utilized for the velocity reconstruction, but the large-scale galaxy modes are instead furnished by SPHEREx, e.g. $P_{gv} \sim \langle \delta_g^{\rm{SPHEREx}} v^{\rm{LSST,CMB}} \rangle$, $P_{gg} \sim \langle \delta_g^{\rm{SPHEREx}} \delta_g^{\rm{SPHEREx}} \rangle $. SPHEREx is an all-sky survey designed to mitigate observational systematics in the measurement of galaxy clustering on large scales~\cite{Dore:2014cca}. SPHEREx will measure spectroscopic redshifts of pre-selected galaxy populations from the all-sky catalogs of the WISE~\cite{WISE}, PanStarrs~\cite{PanSTARRS}, and the DES~\cite{DES} surveys, with its observation pipeline producing a galaxy type, redshift and a redshift uncertainty~\cite{Dore:2014cca}. The strategy of observing multiple galaxy types with different redshift uncertainties lends itself to the use of multi-tracer techniques to mitigate cosmic variance, i.e. internal sample variance cancellation without kSZ, in the inference of primordial non-Gaussianity from large-scale galaxy clustering. SPHEREx will divide its galaxy population, measured over 11 redshift bins out to a maximum redshift of $z_{\rm max} = 4.6$, into 5 different galaxy samples~\cite{Dore:2014cca}, labeled $i$, based on the maximum redshift scatter $\bar{\sigma}_{z,i}$, which ranges from  $\bar{\sigma}_{z,1}=0.003$ to $\bar{\sigma}_{z,5}=0.2$. Details of the number densities, maximum photometric redshift scatters, and linear biases of each galaxy sample are given in Table~\ref{tab:SPHEREx}~\cite{SPHEREx:2014bgr}.\footnote{\url{https://github.com/SPHEREx/Public-products/blob/master/galaxy_density_v28_base_cbe.txt}} The shot noise for each galaxy sample is $N_{g,i}(z) = n_{g,i}^{-1}(z)$, and the photo-z function is $W_i(\mu,k,z) = \exp[-k^2 \mu^2 \sigma_i(z)/2H^2(z)]$, such that the observed galaxy power and cross spectra for two samples $i,j$ are given by
\begin{align}
    P^{\rm{obs}}_{gg,i,j} = b_{g,i}b_{g,j}W_iW_j P_{mm} + N_{g,i} \delta_{ij}\ ,
\end{align}
where $\delta_{ij}$ is the Kronecker delta, and we have suppressed the $(\mu,k,z)$ dependence for brevity.

\begin{table*}[t]
\centering
\caption{Redshift bins $(z_{\mathrm{min}}, z_{\mathrm{max}})$ and corresponding number densities $n_{g,i}$ [$(h/{\rm Mpc})^3$], linear galaxy biases $b_i^{\rm{lin}}$, and (maximum) redshift scatter $\bar{\sigma}_i$ for SPHEREx galaxy samples.}
\label{tab:SPHEREx}
\begin{tabular}{c c|cc|cc|cc|cc|cc}
\hline
 &  & \multicolumn{2}{c|}{Sample 1} & \multicolumn{2}{c|}{Sample 2} & \multicolumn{2}{c|}{Sample 3} & \multicolumn{2}{c|}{Sample 4} & \multicolumn{2}{c}{Sample 5} \\
 &  & \multicolumn{2}{c|}{$\bar{\sigma}_1 = 0.003$} & \multicolumn{2}{c|}{$\bar{\sigma}_2 = 0.01$} & \multicolumn{2}{c|}{$\bar{\sigma}_3 = 0.03$} & \multicolumn{2}{c|}{$\bar{\sigma}_4 = 0.1$} & \multicolumn{2}{c}{$\bar{\sigma}_5 = 0.2$} \\
$z_{\min}$ & $z_{\max}$ & $n_{g,1} \, [h /\rm{Mpc}]^{3}$ & $b_1^{\mathrm{lin}}$ & $n_{g,2} \, [h /\rm{Mpc}]^{3}$ & $b_2^{\mathrm{lin}}$ & $n_{g,3} \, [h /\rm{Mpc}]^{3}$ & $b_3^{\mathrm{lin}}$ & $n_{g,4} \, [h /\rm{Mpc}]^{3}$ & $b_4^{\mathrm{lin}}$ & $n_{g,5} \, [h /\rm{Mpc}]^{3}$ & $b_5^{\mathrm{lin}}$ \\
\hline
0.0 & 0.2 & $9.97\times 10^{-3}$ & 1.3  & $1.23\times 10^{-2}$ & 1.2  & $1.34\times 10^{-2}$ & 1.0  & $2.29\times 10^{-2}$ & 0.98 & $1.49\times 10^{-2}$ & 0.83 \\
0.2 & 0.4 & $4.11\times 10^{-3}$ & 1.5  & $8.56\times 10^{-3}$ & 1.4  & $8.57\times 10^{-3}$ & 1.3  & $1.29\times 10^{-2}$ & 1.3  & $7.52\times 10^{-3}$ & 1.2  \\
0.4 & 0.6 & $5.01\times 10^{-4}$ & 1.8 & $2.82\times 10^{-3}$ & 1.6  & $3.62\times 10^{-3}$ & 1.5  & $5.35\times 10^{-3}$ & 1.4  & $3.27\times 10^{-3}$ & 1.3  \\
0.6 & 0.8 & $7.05\times 10^{-5}$ & 2.3 & $9.37\times 10^{-4}$ & 1.9  & $2.94\times 10^{-3}$ & 1.7  & $4.95\times 10^{-3}$ & 1.5  & $2.50\times 10^{-3}$ & 1.4  \\
0.8 & 1.0 & $3.16\times 10^{-5}$ & 2.1 & $4.30\times 10^{-4}$ & 2.3  & $2.04\times 10^{-3}$ & 1.9  & $4.15\times 10^{-3}$ & 1.7  & $1.83\times 10^{-3}$ & 1.6  \\
1.0 & 1.6 & $1.64\times 10^{-5}$ & 2.7 & $5.00\times 10^{-5}$ & 2.6  & $2.12\times 10^{-4}$ & 2.6  & $7.96\times 10^{-4}$ & 2.2  & $7.34\times 10^{-4}$ & 2.1  \\
1.6 & 2.2 & $3.59\times 10^{-6}$ & 3.6 & $8.03\times 10^{-6}$ & 3.4  & $6.97\times 10^{-6}$ & 3.0  & $7.75\times 10^{-5}$ & 3.6  & $2.53\times 10^{-4}$ & 3.2  \\
2.2 & 2.8 & $8.07\times 10^{-7}$ & 2.3 & $3.83\times 10^{-6}$ & 4.2  & $2.02\times 10^{-6}$ & 3.2  & $7.87\times 10^{-6}$ & 3.7  & $5.41\times 10^{-5}$ & 4.2  \\
2.8 & 3.4 & $1.84\times 10^{-6}$ & 3.2 & $3.28\times 10^{-6}$ & 4.3  & $1.43\times 10^{-6}$ & 3.5  & $2.46\times 10^{-6}$ & 2.7  & $2.99\times 10^{-5}$ & 4.1  \\
3.4 & 4.0 & $1.50\times 10^{-6}$ & 2.7 & $1.07\times 10^{-6}$ & 3.7  & $1.93\times 10^{-6}$ & 4.1  & $1.93\times 10^{-6}$ & 2.9  & $9.41\times 10^{-6}$ & 4.5  \\
4.0 & 4.6 & $1.13\times 10^{-6}$ & 3.8 & $6.79\times 10^{-7}$ & 4.6  & $6.79\times 10^{-7}$ & 5.0  & $1.36\times 10^{-6}$ & 5.0  & $2.04\times 10^{-6}$ & 5.0  \\
\hline
\end{tabular}
\end{table*}

We assume SPHEREx covers a sky fraction $f_{\mathrm{sky}}^{g} = 0.75$ and that the joint area covered by LSST and the CMB survey for velocity reconstruction is $f_{\mathrm{sky}}^{\hat{v}} = 0.3$. We use the same redshift bins $[z_{\rm{min}}, z_{\rm{max}}]$ for LSST and SPHEREx out to $z=2.8$. For simplicity, we assume the entire sky region for velocity reconstruction is contained within the SPHEREx sky fraction. In practice, in our forecasts using LSST velocities and SPHEREx galaxies, this means that there are three distinct regions: 
\begin{enumerate}
    \item the region $f_{\mathrm{sky}}^{\hat{v}} = 0.3$ covered both by LSST and SPHEREx, out to redshift $z=2.8$, for which we have the galaxy measurement and the reconstructed velocities;
    \item the same region $f_{\mathrm{sky}}^{\hat{v}} = 0.3$, but at redshifts $z>2.8$, in which we assume we only have SPHEREx galaxy measurements (as the LSST gold sample is assumed only out to $z_{\rm{max}} \approx 3$);
    \item the remaining sky fraction $f_{\mathrm{sky}}^{g} - f_{\mathrm{sky}}^{\hat{v}}  = 0.45$, for the entire redshift range $z\leq 4.6$, in which we only have  SPHEREx galaxy measurements
\end{enumerate}

Finally, in addition to photometric redshift uncertainties, non-linear bulk-flows smear the (small-scale) BAO features in the observed galaxy power spectrum~\cite{Dore:2014cca}. Following~\cite{Dore:2014cca}, we model this effect by introducing an additional damping factor $f_{BF}(k,z)$ which suppresses small-scale power along and perpendicular to the line of sight
\begin{align}
\label{bulk_flows}
f_{BF}(k,z) = \exp\left(-\frac{1}{2}k^{2}\Sigma_{\perp}^{2}-\frac{1}{2}k^{2}\mu^{2}(\Sigma_{||}^{2}-\Sigma_{\perp}^{2})\right)\ .
\end{align}
where the Lagrangian displacement fields $\Sigma_{\perp}$ and $\Sigma_{||}$ are given as~\cite{Dore:2014cca} 
\begin{eqnarray}
    \Sigma_{\perp}(z) &=& c_{\text{rec}}D(z)\Sigma_{0}\ ,\\
\Sigma_{||}(z) &=& c_{\text{rec}}D(z)(1+f(z))\Sigma_{0}\ ,
\end{eqnarray}
with $D(z)$ and $f(z)$ being the scale-independent linear growth factor and the linear growth rate respectively. We set the parameter $c_{\rm rec} = 0.5$ and $\Sigma_0 = 11\ \text{Mpc/h}$ for $\sigma_{8}=0.8$~\cite{Dore:2014cca}. Marginalising over the parameter $\Sigma_{0}$ around the fiducial value is equivalent to multiplying the Fisher matrix by the square of the exponential factor $f_{BF}$ in Eq.~\eqref{bulk_flows}~\cite{Dore:2014cca}. Although we follow~\cite{Dore:2014cca} in our approach to including bulk flows, we find their inclusion to have a small effect, and they do not have a significant impact on our conclusions.

To carry out our forecasts, we define a Fisher information matrix at each redshift bin $z$. In our forecasts using LSST and the CMB, there is only one sky region. In our SPHEREx-LSST forecasts, we label the sky regions $(f)$. The Fisher matrix in a given $z$ bin and sky region is 
\begin{eqnarray}
F_{\alpha \beta}^{(f)}(z) &=& \frac{V^{(f)}(z)}{2}\int_{-1}^{1} d\mu\int \frac{k^2 dk}{(2\pi)^2}\times \nonumber\\
&&\mathrm{Tr} \left(C^{-1} \frac{dS}{dp_{\alpha}} C^{-1} \frac{dS}{dp_{\beta}} \right)f_{BF}\ , \label{eq:Fisher}
\end{eqnarray}
where $V^{(f)}(z)$ is the survey volume in that redshift bin (which we take to be the cosmological volume in the box geometry, modulated by the corresponding sky fraction $f_{\mathrm{sky}}$), $p_{\gamma}$ is a vector of the marginalized parameters in the forecast, and $C_{ij}=S_{ij}+N_{ij}$ is the covariance matrix, i.e. the sum of the signal and noise matrices,
\begin{align}
    S_{ij} &= \begin{pmatrix}
P_{gg} & P_{gv}\\
P_{v g}  & P_{vv}
\end{pmatrix}\ , \\
N_{ij} &= \begin{pmatrix}
N_g& 0\\
0 & N_{\hat{v}}
\end{pmatrix}\ ,
\end{align}
where the trace runs over the roman indices $(ij)$, which labels the observables, in this case the galaxy and velocity spectra (note we have suppressed these indices in Eq.~(\ref{eq:Fisher}) for notational brevity). The integral in $k$ is bounded by the minimum  and maximum wavenumbers $k_{\mathrm{min}}, k_{\mathrm{max}}$ accessible to the survey. We assume the minimum wavenumber is fixed by the survey volume $k_{\mathrm{min}}(z) =\pi/(V^{1/3}(z)) $, and we take a fixed maximum $k_{\mathrm{max}} = 0.2\,  h/\mathrm{Mpc}$, independent of redshift. We have defined an independent Fisher matrix in each sky region $f$ and redshift bin $z$, so that the total Fisher matrix is the sum 
\begin{align}
    F_{\alpha \beta} = \sum_f \sum_z F_{\alpha \beta}^{(f)}(z)\ ,
\end{align}
where again we note that the sum over sky regions is only present in our SPHEREx-LSST forecasts. Finally, to break parameter degeneracies and improve constraints, we include priors on cosmological parameters from the primary CMB in the form of a Planck-like covariance matrix for the cosmological parameters. We obtain the prior covariance matrix from a mock CMB power spectrum likelihood named `fake\_planck\_realistic' in the publicly available parameter inference package MontePython v3.4~\cite{Audren:2012wb,MontePython3} connected to the Boltzmann code CLASS v3.0.1~\cite{Blas:2011rf,Lesgourgues:2011re,Lesgourgues:2011rh}. This covariance matrix does not encode any CMB bispectrum information and therefore does not provide any information on $\fnlloc$.

In our forecasts, the galaxy bias contains the standard linear bias $b^{\mathrm{lin}}(z)$, the Kaiser term $f\mu^2$ from redshift space distortions \cite{Kaiser:1987qv}, and the scale-dependent contribution in  Eq.~(\ref{eq:bNG_defn}), 
\begin{align}
b_g(\mu, k ,z) = b^{\mathrm{lin}}(z)+\fnlloc \frac{2(b^{\mathrm{{lin}}}(z)-1)\delta_c}{2 k^2 T(k) D(z)/3H_0^2 \Omega_m} + f\mu^2\ ,     
\end{align}
and we marginalize over the galaxy bias parameters $(b^{\mathrm{lin}}_{i}(z), \fnlloc)$ and the kSZ velocity reconstruction bias $b_v(z)$ (where $i$ labels the SPHEREx galaxy sample), with $b^{\mathrm{lin}}_i(z) $ and $b_v(z)$ defined independently in each redshift bin. We assume a fiducial value of $\fnlloc =0$, while the fiducial values for the linear galaxy bias parameters are given in Tables~\ref{tab:LSST} and \ref{tab:SPHEREx}. For cosmological parameters, we marginalize over six $\Lambda\rm{CDM}$ parameters and the sum of the neutrino masses, $(A_s,n_s, \Omega_b h^2, \Omega_c h^2, \theta_s, \tau, \sum m_\nu)$, with corresponding fiducial values $(2.09 \times 10^{-9}, 0.9626, 2.212\times10^{-2}, 1.206 \times 10^{-1}, 1.041126 \times 10^{-2}, 0.0522, 0.06 \,\mathrm{eV})$.

\section{Results \label{sec:results}} 

\begin{figure*}
    \centering
    \includegraphics[width=1.\linewidth]{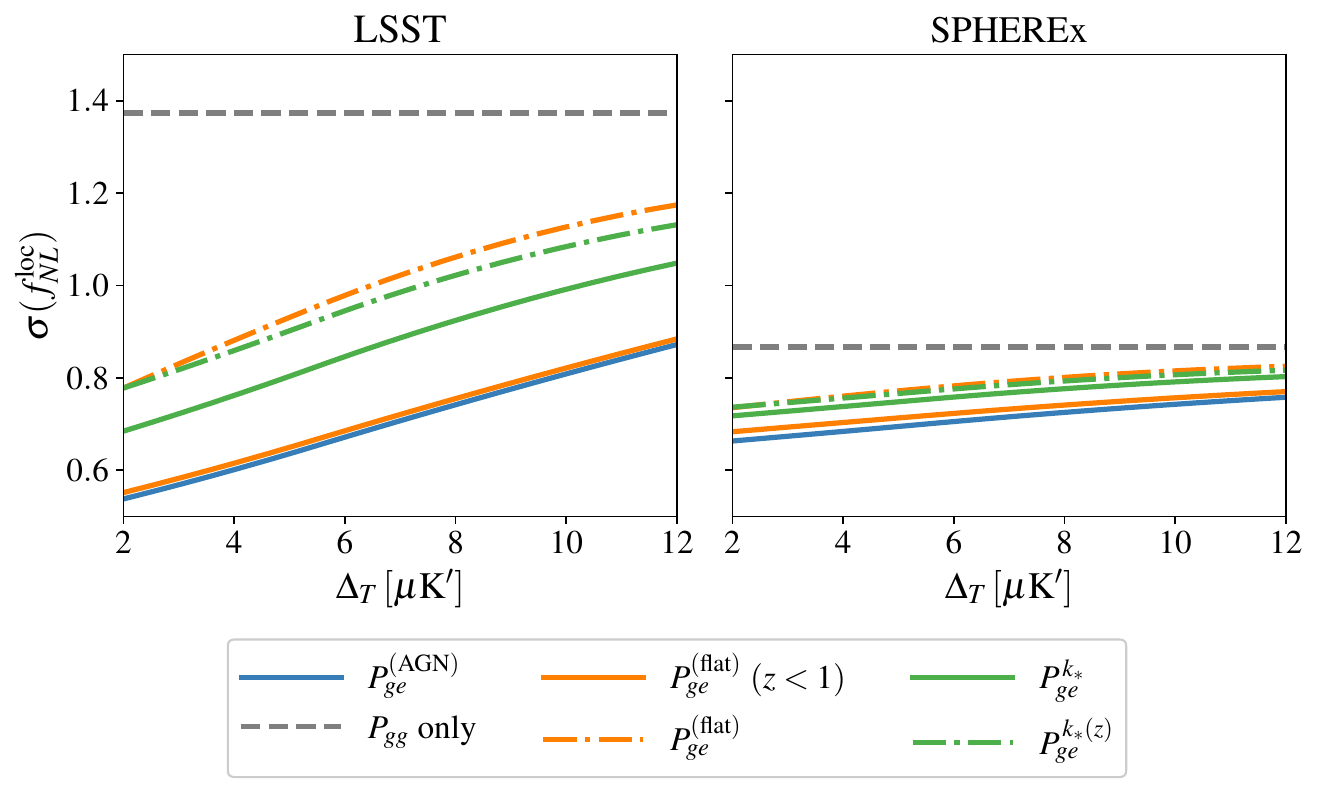}
    \caption{Constraints on $\sigma(\fnlloc)$ from galaxy tomography (gray, dashed lines) and the combination of galaxy and kSZ tomography (solid and dot-dashed lines, in color) from LSST (left panel) and LSST+SPHEREx (right panel) as a function of varying CMB noise $\Delta_T$. The blue curves assume the standard AGN model. The orange lines use a flat suppression of the AGN model, Eq.~(\ref{eq:Pge_suppress_models_flat}), with a suppression factor $A^2(z<1)=0.45, A^2(z>1)=1$ (solid), and $A^2 = 0.45$ for all $z$ (dot-dashed) respectively. The green curves use an exponential suppression of the AGN model on small scales, Eq.(~\ref{eq:Pge_suppress_models_roll}), with a fixed comoving suppression scale $k_* = 2.38 \, \rm{Mpc}^{-1}$ (solid) and a redshift dependent one, $k_*(z) = k_{z_*}(1+z_*)/(1+z)$ (dot-dashed), with $z_* =0.7, k_{z_*} =  2.38\,\rm{Mpc}^{-1}$. Per the main body of the text, the constants $A,k_*$, etc. are chosen to reproduce the value of $b_v = 0.45$ with respect to the AGN model.}
    \label{fig:fNL_DeltaT_LSST_SPHEREx_combined}
\end{figure*}

We present our results for $\fnlloc$ forecasts from galaxy and kSZ tomography from LSST alone and LSST+SPHEREx in the left and right panels of Fig.~\ref{fig:fNL_DeltaT_LSST_SPHEREx_combined} respectively. 

Beginning with our LSST-only forecast, we find $\sigma(\fnlloc) \approx 1.37$ will be achievable with the galaxy measurement alone (gray dashed line), i.e. without information from kSZ velocity reconstruction, in broad agreement with previous studies~\cite{Munchmeyer:2018eey,Adshead:2024paa}. We then include the additional information from kSZ tomography (blue, orange, and green lines), i.e. $P_{gv}$ and $P_{vv}$, while varying both the CMB noise $\Delta_T$ and the assumptions about $P_{ge}$. In blue, we show results assuming the AGN model. As a representative point, for an futuristic, Stage-IV noise level $\Delta_T = 2 \mu K^{\prime}$, we find $\sigma(\fnlloc)\approx 0.53$, again in broad agreement with previous analyses when taking into account slightly different analysis choices~\cite{Munchmeyer:2018eey,Adshead:2024paa}.  In general, we find that the kSZ-driven improvement in $\sigma(\fnlloc)$ will improve moderately as future CMB experiments, e.g. SO, achieve lower noise measurements, yielding an approximately $40$ to $60$ percent reduction in the uncertainty on $\fnlloc$ compared to the galaxy power spectrum in isolation.

However, this picture can change appreciably when we make different assumptions about $P_{ge}$, which we show in the orange and green curves, using the models outlined in Sec.~\ref{subsec:Pge}. In orange, we implement a flat, scale-independent suppression of the AGN model, $P_{ge}^{(\rm{flat})}$ (Eq.(~\ref{eq:Pge_suppress_models_flat})). As detailed in Sec.~\ref{subsec:Pge} we fix the suppression constant $A^2$ to impose agreement with the recent measurements at $z<1$, Refs. \cite{Hotinli:2025tul,Lai:2025qdw}. In the solid orange curve, this suppression is assumed to be present only at $z<1$ (i.e. we set $A=1$ at $z>1$, recovering the AGN model), while in the dot-dashed orange curve assumes $A^2 =0.45$ over the entire redshift range. In green, we relax the assumption of scale-independent suppression of $P_{ge}$, and instead implement the exponential suppression of Eq.~(\ref{eq:Pge_suppress_models_roll}). The solid curve takes a constant comoving suppression scale, $k_* = \rm{const} = 2.38/\rm{Mpc}$, which is chosen to recover the $b_v = 0.45$ of Ref.~\cite{Hotinli:2025tul} when matching the survey specifications and redshift range of their analysis. By contrast, the dot-dashed green curve assumes the comoving suppression scale is redshift dependent, $k_*(z) = k_{z_*} (1+z_*)/(1+z)$ (i.e. a fixed physical wavenumber $k_{\rm{phys}}$), where again the constants are chosen to reproduce $b_v(z_*=0.7)=0.45$, $k_{z_*} = k_* = 2.38\, \rm{Mpc}^{-1}$. 

\begin{figure}
    \centering
    \includegraphics[width=1.\linewidth]{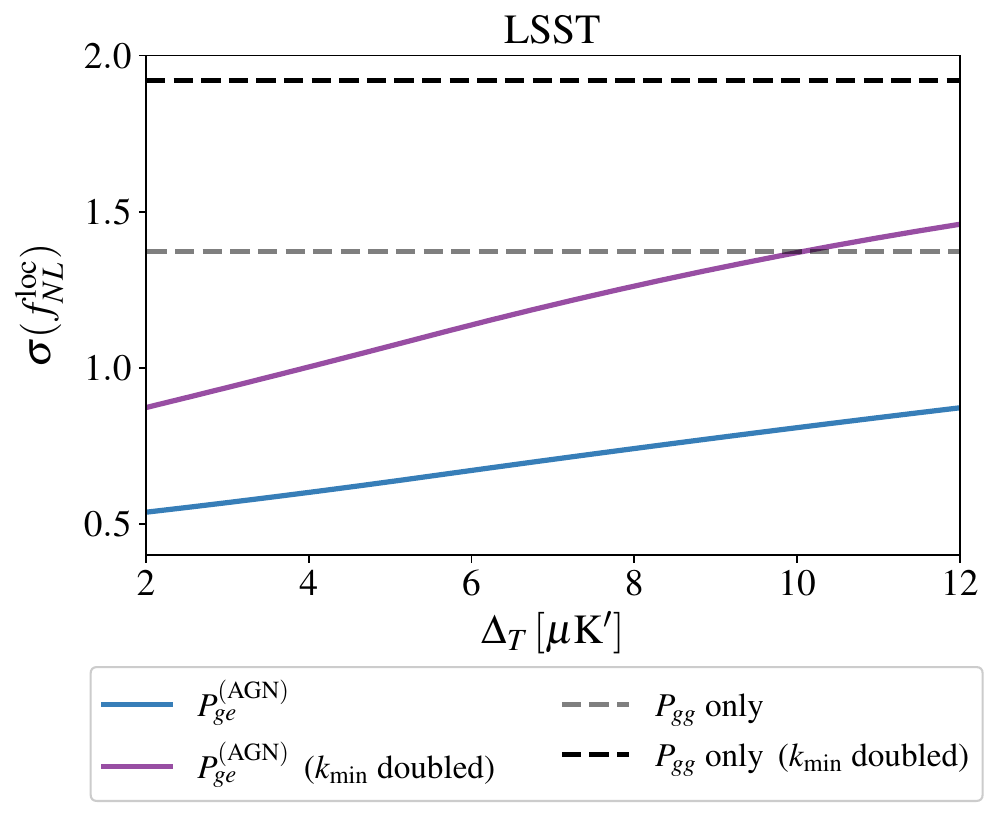}
    \caption{Impact of large-scale modes (i.e., $k_{\rm{min}}$) on $\sigma(\fnlloc)$ from galaxy tomography (dashed lines) and the combination of galaxy and kSZ tomography (solid lines), assuming an LSST-like survey (Table~\ref{tab:LSST}). Our fiducial results for the galaxy power spectrum alone and the combination of galaxy and kSZ tomography (assuming the AGN model), with $k_{\rm{min}} =\pi/V^{1/3}(z)$, are shown again in dashed gray and solid blue, respectively. The impact on these constraints from removing large-scale information, by doubling the minimum wavenumber in the Fisher matrix, is shown in black dashed (galaxies alone) and solid purple (galaxies and kSZ) respectively.}
    \label{fig:kmin_fNL}
\end{figure}

As shown earlier in Fig.~\ref{fig:Nv_Pge_compare}, the orange and green curves all assume a galaxy-electron spectrum that is suppressed compared to the AGN model, which yields a larger projected reconstruction noise and hence worse $\fnlloc$ constraints. Focusing first on the orange curves, the flat suppression of $P_{ge}$ only significantly affects the $\fnlloc$ constraints if it is present at high  ($z>1$) redshifts. The reason for this is that high redshift information is crucial for $\fnlloc$ constraints because it provides access to the largest scale modes, where the scale-dependent bias signal $\Delta b\sim 1/k^2$ is strongest and the efficacy of sample-variance cancellation is greater (see Fig.~\ref{fig:kmin_fNL} for an example of how the loss of large-scale information worsens $\fnlloc$ constraints). Increasing the reconstruction noise at only low redshifts therefore has little effect on $\sigma(\fnlloc)$, as kSZ is simply not providing very much additional constraining power for $\fnlloc$ at those redshifts. On the other hand, as the dot-dashed orange curve indicates, if the suppression of small-scale galaxy-electron power observed at $z<1$ also persists at higher redshifts, kSZ tomography will offer less constraining power compared to the predictions made with the AGN model, in this example leading to a $\approx \!30-40\%$ increase in the measurement uncertainty $\sigma(\fnlloc)$.

The scale-dependent, exponential suppression of $P_{ge}$ (green curves) yields results that differ quantitatively, but the upshot is the same. As we discussed earlier in Sec.~\ref{subsec:Pge}, the kSZ signal peaks near $\ell \sim 4000$, meaning the wavenumbers $k_s = \ell/\chi(z)$ contributing to the reconstruction move to smaller wavenumbers (larger scales) at higher redshifts. Consequently, the suppression for  $k_*=\rm{const}$ is more significant at lower redshifts than at higher redshifts, which can also be seen in the noise curves in Fig.~\ref{fig:Nv_Pge_compare}. Compared to the assumption of a flat suppression at all redshifts (orange dot-dashed), the assumption of a fixed comoving $k_*$ comparatively reduces (increases) the reconstruction noise for redshifts above (below) $z=0.7$. Therefore, due to the  importance of high-redshift information in the sample variance cancellation approach, the fixed $k_*$ model yields a slightly smaller increase in $\sigma(\fnlloc)$ of approximately $20-30\%$ over the AGN prediction (compared to the $30-40\%$ when assuming a flat suppression over all redshifts).

Finally, in the dot-dashed green curve, we show results assuming the exponential suppression of the AGN model with a redshift dependent comoving suppression scale, $k_*(z) = k_{z_*}(1+z_*)/(1+z)$. As discussed in Sec.~\ref{subsec:Pge}, this simple redshift dependence, $\propto 1/(1+z)$, corresponds to a redshift-independent \textit{physical} wavenumber $k_{*,\rm{phys}} = \rm{const}$, i.e. a characteristic length scale $r_*$ in the system below which galaxies and free electrons become decorrelated. This choice also moves the suppression scale to smaller wavenumbers at higher redshifts, yielding $P_{ge}(z)$ that is comparatively lower for $z>z_*$ compared to the fixed $k_*$ model shown in solid green, and hence slightly weaker $\fnlloc$ constraints, again due to the importance of higher redshift information. In this case, the LPnG measurement uncertainty is again roughly $30-40\%$ larger than when assuming the AGN model, similar to assuming a flat suppression at all redshifts.

Although the details differ, a qualitatively similar picture emerges when we instead consider cross-correlating reconstructed velocities from LSST with large-scale galaxy modes from SPHEREx, shown in the right panel of Fig.~\ref{fig:fNL_DeltaT_LSST_SPHEREx_combined}. While the science case is roughly the same, there are some key differences when utilizing SPHEREx. SPHEREx is (in part) an $\fnlloc$ machine, designed specifically to achieve $\sigma(\fnlloc) \lesssim 1$ via large-scale galaxy measurements across several distinct samples~\cite{SPHEREx:2014bgr}, exploiting sample variance cancellation in galaxy clustering measurements alone. For this reason, the $\fnlloc$ constraint from SPHEREx alone is significantly stronger than LSST can achieve when considering galaxy power spectrum measurements in isolation. Here, we find $\sigma(\fnlloc) \approx 0.87$, in excellent agreement with previous analyses (see e.g.~\cite{Dore:2014cca}) despite minor differences in analysis choices.\footnote{We note that our SPHEREx forecasts in Fig.~\ref{fig:fNL_DeltaT_LSST_SPHEREx_combined} are slightly conservative due to the fact that splitting the sky causes some loss of large-scale information (namely $k_{\rm{min}}$ is greater by a factor $(V/V')^{1/3} \approx 2^{1/3}$, where $V,V'$ are the survey volumes in the unsplit and split sky treatments respectively). In practice, we find this effect is modest, yielding $\lesssim10\%$ higher uncertainty $\sigma(\fnlloc)$. We find sub-percent agreement in $\sigma(\fnlloc)$ with Ref.~\cite{SPHEREx:2014bgr} when using the same (unsplit) sky fraction and redshift independent $k_{\mathrm{min}}= 10^{-3} h/\,\rm{Mpc}$. Our primary focus, namely the effect of $P_{ge}$ on $\fnlloc$ constraints from kSZ tomography, is insensitive to these details.} Owing to the internal sample variance cancellation, the role of kSZ tomography is less distinctive when combining with SPHEREx measurements, and so the reduction in $\sigma(\fnlloc)$ made by including information from the velocity field are comparatively smaller than in the LSST-only analysis. For example, for the reference AGN model and $\Delta_T = 2 \mu K'$, our forecast for the combination of SPHEREx galaxy and LSST kSZ measurements yields $\sigma(\fnlloc) \approx 0.66$, similar to but slightly higher than from the combination of LSST galaxies and velocities under the same assumptions. Although our forecasts for LSST take the gold sample only out to $z<3$, and SPHEREx goes to higher redshift $z_{\rm{max}}=4.6$, LSST benefits from significantly higher number densities, which decreases the kSZ reconstruction noise and which previous studies have shown significantly improves the power of sample variance cancellation approaches to probing $\fnlloc$~\cite{Munchmeyer:2018eey}. 

Despite these mild differences coming from the galaxy survey details, the conclusions are essentially the same. As we modify $P_{ge}$ and vary the CMB noise, the same trends emerge in $\sigma(\fnlloc)$. The  impact of suppressing $P_{ge}$ for only $z<1$ is comparatively slightly greater than in the LSST-only case, indicating that low redshift information is comparatively slightly more important when combining the kSZ velocities with SPHEREx galaxies, but the effect is small.  The general trends remain the same, reinforcing the importance of high-redshift information. Owing to the internal sample variance cancellation of SPHEREx, the results here are overall less sensitive to the varying assumptions about the galaxy-electron spectrum; the inclusion of kSZ tomography is less impactful, but somewhat more robust to modeling ambiguities in $P_{ge}$ (e.g. yielding $\mathcal{O}(10\%)$ degradation in $\sigma(\fnlloc)$ with respect to the AGN model, rather than $20-40\%$ in the case of LSST). On the whole, the conclusions for both surveys are very similar: for the scenarios we have examined here --- which are broadly consistent with present measurements, but not exhaustive --- kSZ tomography will remain a useful, but perhaps less powerful tool for constraining local primordial non-Gaussianity than previously estimated with forecasts utilizing the AGN model.

\section{Discussion \label{sec:discussion}}
In this paper we have explored how modifying assumptions of the underlying galaxy-electron spectrum impacts prospects for constraining local type primordial non-Gaussianity with kSZ tomography with current and forthcoming cosmological surveys. Specifically, we have explored the possibility that significant suppression of $P_{ge}$ compared to the AGN model, implied by $b_v<1$ measurements from recent velocity reconstructions at $z \lesssim 1$~\cite{Hotinli:2025tul,Lai:2025qdw}, persists at higher redshifts. Our main conclusion, represented in Fig.~\ref{fig:fNL_DeltaT_LSST_SPHEREx_combined}, is that such scenarios can significantly reduce the constraining power of kSZ tomography, as evidenced by the increase in the measurement error $\sigma(\fnlloc)$, depending on the details of the suppression. Our analysis indicates that, among the range of effects we have considered here, the critical feature is the average amplitude of $P_{ge}$ on small scales (around $k_S \approx 4000/\chi(z)$) at redshifts $z\gtrsim 1$.

Our approach here has been to broadly and simply extrapolate the potential implications of recent measurements~\cite{Hotinli:2025tul,Lai:2025qdw}. We have explored simple scale- and redshift-dependent suppression of the galaxy-electron spectrum, calibrated to the AGN model at $z_* =0.7$ and extrapolated to higher redshifts, where a significant fraction of the information for $\fnlloc$ constraints comes from due to access to larger scale modes, for which the scale-dependent bias signal is stronger. However, the space of possibilities is large, and the scenarios we have considered here are by no means exhaustive. Future measurements could reveal that the overestimation of $P_{ge}$ is even more severe at higher redshifts (i.e. beyond the range $z\lesssim 1$ probed by recent velocity reconstruction measurements~\cite{Hotinli:2025tul, Lai:2025qdw}), which would further reduce the efficacy of kSZ tomography vis-à-vis previous expectations. On the other hand, the situation could equally well turn out to be far more favorable, and the gas profiles commonly adopted in the literature could be more accurate at higher redshifts --- or even underestimates --- in which case future prospects for kSZ tomography will be brighter. The critical quantity is the galaxy-electron power on small scales (especially at higher redshifts for LPnG constraints), as this controls the strength of the kSZ signal. The detailed response of the velocity reconstruction noise, and measurement uncertainty on cosmological parameters, are sensitive to how exactly that power varies as a function of scale and cosmic history. Velocity reconstructions at $z \gtrsim 1$ will illuminate the situation.

The role of kSZ tomography as a probe of cosmology, at least in the context of constraints on $\fnlloc$ with forthcoming surveys, is perhaps not clear, and potentially weaker than previously expected. Whether or not that will materialize remains to be seen. Another perspective, however, is that recent kSZ measurements \cite{Hadzhiyska:2024qsl,Hotinli:2025tul,Lai:2025qdw}, in synergy with other probes, e.g. thermal SZ, weak lensing, CMB lensing, X-ray observations \cite{Hadzhiyska:2025mvt,Siegel:2025frt,Dalal:2025ssl,ACT:2025llb}, have uncovered interesting, and perhaps surprising results about the galaxy-electron connection on small scales, at least compared to profiles often adopted in contemporary kSZ analyses. In this way, we imagine several promising future directions for this work. For example, it would be interesting to explore models of the galaxy-electron spectrum in greater detail. One possibility would be to analyze dependence of the suppression scale $k_*$ as a function of halo mass and redshift (in contrast to just marginalizing over $b_v$), which is a direction we plan to investigate in the future. Another interesting approach would be to leverage results from simulations including baryonic feedback in massive halos (e.g. \cite{Springel:2017tpz,Schaye:2023jqv}). In all, future velocity reconstructions, especially at higher redshifts, may surprise us yet further, and regardless of what they imply about the cosmological utility of kSZ tomography, they will serve as crucial probes of small-scale astrophysics and feedback processes. From this point of view, the emerging program of kSZ velocity reconstruction is a robust avenue for discovery.

\section{Acknowledgments}
AJT is supported by the United States Department of Energy, DE-SC0015655. CS acknowledges support from the David and Lucile Packard Foundation; NASA grant 22-ROMAN11-0011, via a subaward from the Jet Propulsion Laboratory; and the SPHEREx project under a contract from the NASA/Goddard Space Flight Center to the California Institute of Technology. GH is supported by Brand and Monica Fortner.

\bibliography{main}

\end{document}